\documentclass{iau}
\usepackage{graphicx}

\usepackage{gensymb}
\usepackage{hyperref}

\begin{document}

\lefttitle{J. D. Sakowska et al.}
\righttitle{Stellar streams around dwarf galaxies are observationaly rare}

\jnlPage{1}{7}
\jnlDoiYr{2026}
\doival{10.1017/xxxxx}

\aopheadtitle{Proceedings of IAU Symposium 403}
\editors{D. Martínez-Delgado,  eds.}

\title{Stellar streams around dwarf galaxies are observationally rare in the local Universe}



\author{
Joanna D. Sakowska$^1$, David~Mart\'inez-Delgado$^{2,3}$, Sarah Pearson$^{4,5}$, Francisco J. Riquel-Castilla$^6$, Tjitske K. Starkenburg$^{7,8}$, Giuseppe Donatiello$^9$, Alis Deason$^{10}$, Denis Erkal$^{11}$, Ethan D. Taylor$^{11}$}

\affiliation{
$^{1}$ Instituto de Astrofísica de Andalucía (IAA-CSIC), Glorieta de la Astronom\'\i a,  E-18008 Granada, Spain\\
$^{2}$ Centro de Estudios de F\'isica del Cosmos de Arag\'on (CEFCA), Unidad Asociada al CSIC, Plaza San Juan 1, 44001 Teruel, Spain\\
$^{3}$ ARAID Foundation, Avda. de Ranillas, 1-D, E-50018 Zaragoza, Spain\\
$^{4}$ Niels Bohr International Academy $\&$ DARK, Niels Bohr Institute, University of Copenhagen, Blegdamsvej 17, 2100 Copenhagen, Denmark\\
$^{5}$ DTU Space, Technical University of Denmark, Elektrovej 327, DK- 2800 Kongens Lyngby, Denmark
$^{6}$ Facultad de Físicas, Universidad de Sevilla, Avda. Reina Mercedes s/n, Campus Reina Mercedes, E-41012 Seville, Spain\\
$^{7}$ CIERA and Department of Physics and Astronomy, Northwestern University, 1800 Sherman Ave, Evanston, IL 60201, USA \\
$^{8}$ NSF–Simons AI Institute for the Sky (SkAI), 172 E. Chestnut St., Chicago, IL 60611, USA\\
$^{9}$UAI -- Unione Astrofili Italiani /P.I. Sezione Nazionale di Ricerca Profondo Cielo, 72024 Oria, Italy \\
$^{10}$ Institute for Computational Cosmology, Department of Physics, Durham University, South Road, Durham DH1 3LE, UK\\
$^{11}$ Department of Physics, University of Surrey, Guildford GU2 7XH, UK \\
}

\begin{abstract}
The frequency and properties of stellar streams around dwarf galaxies remain observationally under-explored. Building a statistically significant sample is challenging because dwarf galaxies are smaller, fainter, and far more numerous than massive galaxies. We present the first results from a systematic survey for stellar streams around dwarf galaxies. We first introduce a classification metric for satellite accretion features (streams, shells, and asymmetric stellar haloes) using DECam imaging from the DES and DECaLS surveys. Applying this metric to the DES footprint, we find a low observed frequency of accretion features (5.1$\%$) and no bona fide stellar streams (with only one identified in DECaLS). We consider whether the apparent excess of shells relative to streams can be explained by observational biases. Our results indicate that stellar streams are observationally rare in the local Universe, motivating further theoretical work to determine whether this reflects genuine differences in merger rates across galaxy mass regimes.}
\end{abstract}

\maketitle

Within the $\Lambda$CDM paradigm, galaxies assemble hierarchically through the accretion and merger of smaller systems (e.g., \citealt{WR78}). Consequently, dwarf galaxies ($M_\star < 10^{9.5}M_\odot$) are expected either to host surviving satellites, or to have experienced satellite accretion during their evolution (\citealt{W15, D17}). Low surface brightness studies of the nearby Magellanic Clouds provide compelling evidence for this process: the Small Magellanic Cloud, a $\sim$1:10 mass-ratio satellite of the Large Magellanic Cloud (LMC), shows tidal debris and stellar structures linked to interactions with LMC (e.g., \citealt{B16, PM20, Culli2023, PM24, Sakowska24}). Further, there are indications that the Clouds may host a population of their own ultra-faint satellites
(e.g., \citealt{Bechtol2015, Koposov2015, Drlicawagner2015, Martin2015, Koposov2018, Torrealba2018, Cerny2023, Pace2025, mg26}). Extending such studies to statistically large samples of dwarf galaxies is crucial for stress-testing $\Lambda$CDM, particularly because small-structure formation remains among the strongest observational tests of cosmological paradigms (see \citealt{Bullock2017} and references therein).

Stellar streams provide a long-lived record of satellite accretion. Even after the satellite has been fully disrupted, the properties of its stream describe its progenitor's properties (orbital history, mass, etc.). For example, the shape of the stream (long stream versus shell) is linked to the satellite's infall trajectory (radial versus circular/polar). Comparing the observed frequency and morphology of stellar streams with predictions from cosmological simulations therefore offers a direct test of hierarchical galaxy formation. While over one hundred stellar streams have now been identified around Milky Way-mass galaxies (\citealt{MD2023, MC2023, MC2024, Sola2025}) and a subset have been compared against predictions from cosmological predictions (\citealt{MC2025}), equivalent studies are scarce for dwarf galaxies. The work presented here, and subsequently published in \cite{Sakowska2026}, is a step towards this goal in the low-mass regime.

\begin{figure}
\begin{center}
\includegraphics[width=1\textwidth]{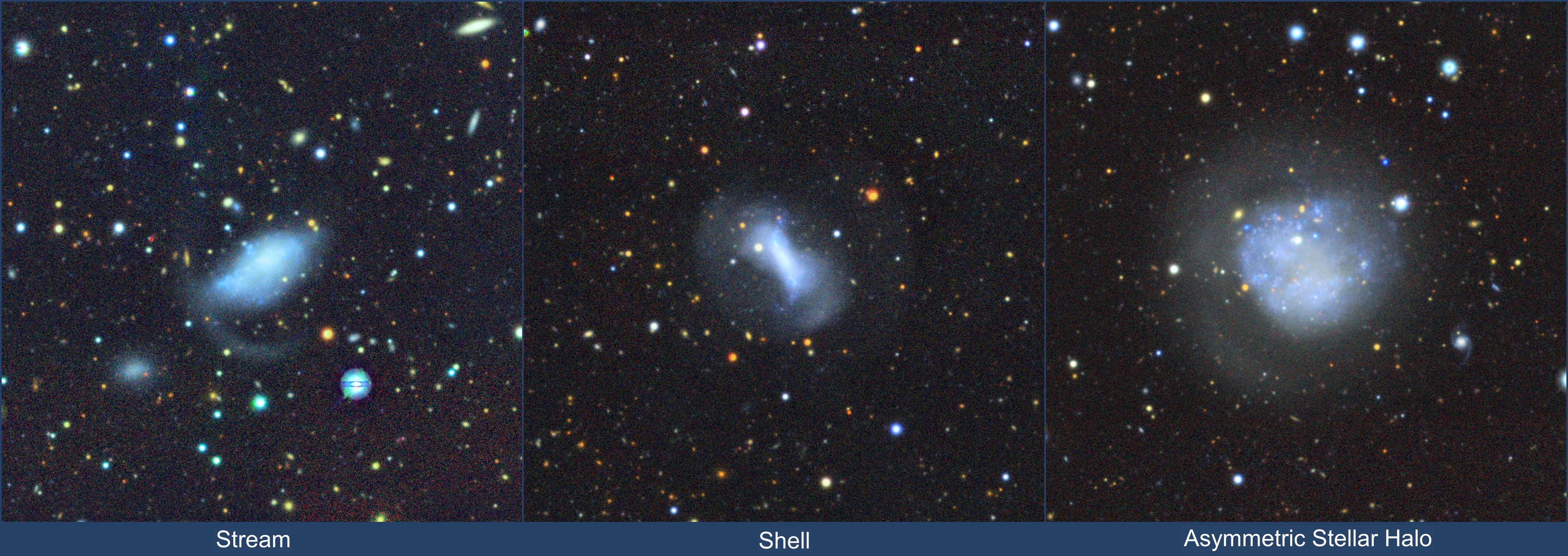}
\caption{Our classification of stellar streams in dwarfs from our visual inspection of DES and DECaLS survey images: stream (ESO 508-509), shell (PGC 40604) and asymmetric stellar halo (PGC 46382).
}\label{fig:morphology}
\end{center}
\end{figure}

\begin{figure}
\begin{center}
\includegraphics[width=1\textwidth]{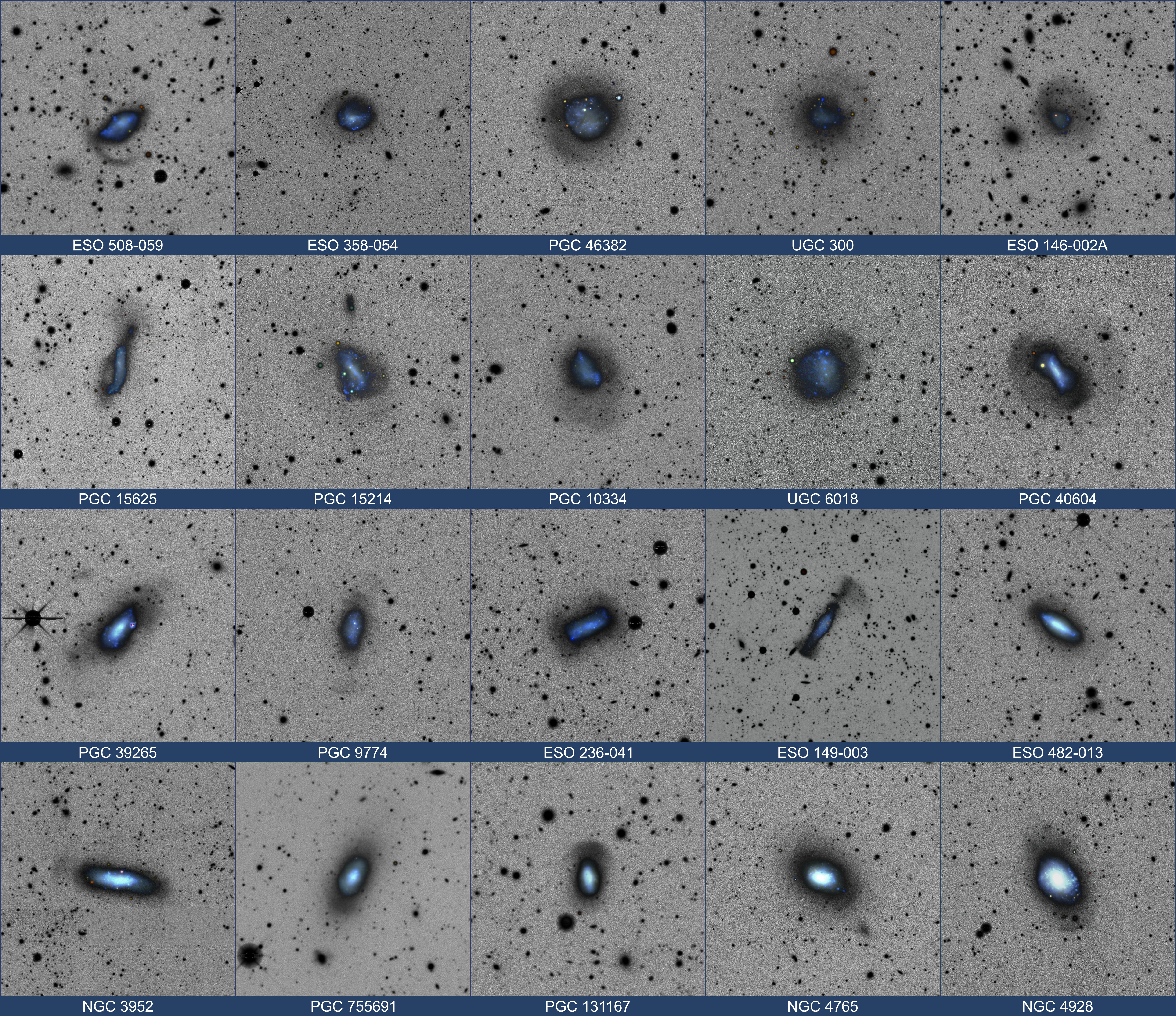}
\caption{Examples of accretion features around dwarfs classified using the metric in Figure \ref{fig:morphology}: we find a stream (upper left, ESO 508-509), asymmetric stellar halos (ESO 358-054 - UGC 6018) and shells (PGC 40604 - NGC 4928). We found only one stream morphology, underscoring their observational rarity in the local Universe. }\label{fig:mosaic}
\end{center}
\end{figure}

We have quantified the frequency of stellar streams around dwarf galaxies through a systematic, visual inspection of isolated dwarfs within the DES footprint. Given limited cases of streams and shells around dwarf galaxies are known (e.g., \citealt{MD12, Annibali2016, paudel18, kado-fong2020, annibali2022, pascale2022, pascale2024, sacchi2024, Fielder2025}), our first step consisted of converging on a morphological classification metric for classifying stellar stream phenomena in the low-mass regime across DECam images. To ensure the metric represents a sufficiently large sample of isolated dwarf galaxies, we visually inspected the DECaLS footprint in addition to DES. In Figure \ref{fig:morphology} we present the classification metric: stellar streams, which are not part of the disc (ESO 508-059); discernable shells (PGC 40606) and asymmetric stellar halos (PGC 46382). In Figure \ref{fig:mosaic} we compile 20 of the most spectacular examples of stellar streams found (DES and DECaLS). Similar to Fig. \ref{fig:morphology}, the streams are ordered by apparent morphology: stellar stream (ESO 508-059), asymmetric stellar halo (ESO 358-054 to UGC 6018) and shell (IC 700 to NGC 4928). 17 of these 20 cases are new discoveries (see appendix A of \citealt{Sakowska2026} for a description of each case).

Applying this classification metric to the DES sample, we measure the observable frequency of streams, shells and asymmetric stellar halos as 5.1$\%$ for isolated dwarf galaxies. This result agrees with independent in preparation estimates on the observability of tidal debris around LMC-sized galaxies ($\sim$4 - 11$\%$ for DECam, \citealt{Starkenburg}). By comparison, using the same DES data and surface brightness (SB) limits, \cite{MC2024} found that $9.1\% \pm 1.1\%$ of massive galaxies showed stellar streams. While this could point towards genuine differences for merger rates and accretion debris morphologies across mass regimes, potential observational biases complicate direct comparisons.

One of the main outcomes of our study is the observed lack of \textit{bona-fide} long streams around dwarf galaxies (only ESO 508-059 in DECaLS), implying an observational detection bias towards shell-like streams. On the one hand, dwarf major mergers may indeed preferentially form shells; for massive galaxies, dynamical friction can radialise orbits when the accreted satellite is relatively massive (\citealt{Amorisco2017}). If the same process operates in the dwarf regime, major mergers may naturally produce shells more frequently than streams. Because major mergers also contribute more stellar mass to the debris, the resulting shells should remain brighter and detectable for longer (before they phase mix into asymmetric stellar halos) than the lower SB streams produced by minor mergers. On the other hand, stream orientation could contribute towards morphological and detectability bias. Given our classifications rely on the brightest pieces of tidal debris above the survey's SB limit, the detectability and apparent morphology of the stream should be affected by its viewing angle. For example, a stellar stream orientated edge-on along the line of sight will appear brighter; in certain cases, an edge-on stellar stream may also appear shell-like. These effects could bias classifications towards shell-like morphologies.

Overall, our results demonstrate that streams around dwarf galaxies are observationally rare. Following the discussion above, we regard 5.1$\%$ as an upper limit on the observable fraction of dwarf galaxy streams in the DES sample. The considerably low SB of dwarf galaxy streams, together with the limited theoretical modelling of dwarf-dwarf mergers, has made it challenging to unambiguously distinguish accretion from morphological features in dwarfs (see appendix B of \citealt{Sakowska2026}). An extensive library of hydrodynamical $N$-body simulations spanning the full dwarf-dwarf merger sequence, while accounting for SB limits and viewing angle effects, would help constrain our measurement. Deep, wide-field surveys (such as Euclid or the LSST), combined with upcoming cosmological predictions on the detectability of dwarf galaxy streams, should verify whether there are true differences for merger rates across mass regimes in the local Universe.


\begin{thebibliography}{}

\bibitem[\protect\citeauthoryear{Amorisco}{2017}]{Amorisco2017} Amorisco N.~C., 2017, MNRAS, 464, 2882. 

\bibitem[\protect\citeauthoryear{Annibali et al.}{2016}]{Annibali2016} Annibali F., Nipoti C., Ciotti L., Tosi M., Aloisi A., Bellazzini M., Cignoni M., et al., 2016, ApJL, 826, L27. 

\bibitem[\protect\citeauthoryear{Annibali et al.}{2022}]{annibali2022} Annibali F., Bacchini C., Iorio G., Bellazzini M., Pascale R., Beccari G., Cignoni M., et al., 2022, MNRAS, 512, 1781. 

\bibitem[\protect\citeauthoryear{Bechtol et al.}{2015}]{Bechtol2015} Bechtol K., Drlica-Wagner A., Balbinot E., Pieres A., Simon J.~D., Yanny B., Santiago B., et al., 2015, ApJ, 807, 50.

\bibitem[\protect\citeauthoryear{Besla et al.}{2016}]{B16} Besla G., Mart{\'\i}nez-Delgado D., van der Marel R.~P., Beletsky Y., Seibert M., Schlafly E.~F., Grebel E.~K., et al., 2016, ApJ, 825, 20.

\bibitem[\protect\citeauthoryear{Bullock \& Boylan-Kolchin}{2017}]{Bullock2017} Bullock J.~S., Boylan-Kolchin M., 2017, ARA\&A, 55, 343.

\bibitem[\protect\citeauthoryear{Cerny et al.}{2023}]{Cerny2023} Cerny W., Mart{\'\i}nez-V{\'a}zquez C.~E., Drlica-Wagner A., Pace A.~B., Mutlu-Pakdil B., Li T.~S., Riley A.~H., et al., 2023, ApJ, 953, 1. 
\bibitem[\protect\citeauthoryear{Cullinane et al.}{2023}]{Culli2023} Cullinane L.~R., Mackey A.~D., Da Costa G.~S., Koposov S.~E., Erkal D., 2023, MNRAS, 518, L25. 
\bibitem[\protect\citeauthoryear{Dooley et al.}{2017}]{D17} Dooley G.~A., Peter A.~H.~G., Carlin J.~L., Frebel A., Bechtol K., Willman B., 2017, MNRAS, 472, 1060. 

\bibitem[\protect\citeauthoryear{Drlica-Wagner et al.}{2015}]{Drlicawagner2015} Drlica-Wagner A., Bechtol K., Rykoff E.~S., Luque E., Queiroz A., Mao Y.-Y., Wechsler R.~H., et al., 2015, ApJ, 813, 109.

\bibitem[\protect\citeauthoryear{Fielder et al.}{2025}]{Fielder2025} Fielder C.~E., Sand D.~J., Jones M.~G., Crnojevi{\'c} D., Drlica-Wagner A., Bennet P., Carlin J.~L., et al., 2025, ApJL, 982, L41. 
\bibitem[\protect\citeauthoryear{Kado-Fong et al.}{2020}]{kado-fong2020} Kado-Fong E., Greene J.~E., Greco J.~P., Beaton R., Goulding A.~D., Johnson S.~D., Komiyama Y., 2020, AJ, 159, 103. 
\bibitem[\protect\citeauthoryear{Koposov et al.}{2015}]{Koposov2015} Koposov S.~E., Belokurov V., Torrealba G., Evans N.~W., 2015, ApJ, 805, 130. 
\bibitem[\protect\citeauthoryear{Koposov et al.}{2018}]{Koposov2018} Koposov S.~E., Walker M.~G., Belokurov V., Casey A.~R., Geringer-Sameth A., Mackey D., Da Costa G., et al., 2018, MNRAS, 479, 5343.

\bibitem[\protect\citeauthoryear{Martin et al.}{2015}]{Martin2015} Martin N.~F., Nidever D.~L., Besla G., Olsen K., Walker A.~R., Vivas A.~K., Gruendl R.~A., et al., 2015, ApJL, 804, L5.

\bibitem[\protect\citeauthoryear{Mart{\'\i}nez-Delgado et al.}{2012}]{MD12} Mart{\'\i}nez-Delgado D., Romanowsky A.~J., Gabany R.~J., Annibali F., Arnold J.~A., Fliri J., Zibetti S., et al., 2012, ApJL, 748, L24. 
\bibitem[\protect\citeauthoryear{Mart{\'\i}nez-Delgado et al.}{2023}]{MD2023} Mart{\'\i}nez-Delgado D., Cooper A.~P., Rom{\'a}n J., Pillepich A., Erkal D., Pearson S., Moustakas J., et al., 2023, A\&A, 671, A141. 

\bibitem[\protect\citeauthoryear{Mart{\'\i}nez-Garc{\'\i}a et al.}{2026}]{mg26} Mart{\'\i}nez-Garc{\'\i}a A.~M., del Pino A., van der Marel R.~P., Battaglia G., {\L}okas E.~L., Vitral E., McKinnon K.~A., et al., 2026, arXiv, arXiv:2606.13787. 

\bibitem[\protect\citeauthoryear{Massana et al.}{2020}]{PM20} Massana P., No{\"e}l N.~E.~D., Nidever D.~L., Erkal D., de Boer T.~J.~L., Choi Y., Majewski S.~R., et al., 2020, MNRAS, 498, 1034. 

\bibitem[\protect\citeauthoryear{Massana, Nidever, \& Olsen}{2024}]{PM24} Massana P., Nidever D.~L., Olsen K., 2024, MNRAS, 527, 8706. 

\bibitem[\protect\citeauthoryear{Mir{\'o}-Carretero et al.}{2023}]{MC2023} Mir{\'o}-Carretero J., Mart{\'\i}nez-Delgado D., Farr{\`a}s-Aloy S., G{\'o}mez-Flechoso M.~A., Cooper A., Roca-F{\`a}brega S., Kuijken K., et al., 2023, A\&A, 669, L13.

\bibitem[\protect\citeauthoryear{Mir{\'o}-Carretero et al.}{2024}]{MC2024} Mir{\'o}-Carretero J., Mart{\'\i}nez-Delgado D., G{\'o}mez-Flechoso M.~A., Cooper A., Akhlaghi M., Donatiello G., Kuijken K., et al., 2024, A\&A, 691, A196. 

\bibitem[\protect\citeauthoryear{Mir{\'o}-Carretero et al.}{2025}]{MC2025} Mir{\'o}-Carretero J., G{\'o}mez-Flechoso M.~A., Mart{\'\i}nez-Delgado D., Cooper A.~P., Roca-F{\`a}brega S., Akhlaghi M., Pillepich A., et al., 2025, A\&A, 700, A176.

\bibitem[\protect\citeauthoryear{Pace et al.}{2025}]{Pace2025} Pace A.~B., Li T.~S., Ji A.~P., Simon J.~D., Cerny W., Senkevich A.~M., Drlica-Wagner A., et al., 2025, OJAp, 8, 112. 

\bibitem[\protect\citeauthoryear{Pascale et al.}{2022}]{pascale2022} Pascale R., Annibali F., Tosi M., Marinacci F., Nipoti C., Bellazzini M., Romano D., et al., 2022, MNRAS, 509, 2940. 

\bibitem[\protect\citeauthoryear{Pascale et al.}{2024}]{pascale2024} Pascale R., Annibali F., Tosi M., Nipoti C., Marinacci F., Bellazzini M., Cannon J.~M., et al., 2024, A\&A, 688, A144. 

\bibitem[\protect\citeauthoryear{Paudel et al.}{2018}]{paudel18} Paudel S., Smith R., Yoon S.~J., Calder{\'o}n-Castillo P., Duc P.-A., 2018, ApJS, 237, 36. 

\bibitem[\protect\citeauthoryear{Sacchi et al.}{2024}]{sacchi2024} Sacchi E., Bellazzini M., Annibali F., Tosi M., Beccari G., Cannon J.~M., Hunter L.~C., et al., 2024, A\&A, 691, A65. 

\bibitem[\protect\citeauthoryear{Sakowska et al.}{2024}]{Sakowska24} Sakowska J.~D., No{\"e}l N.~E.~D., Ruiz-Lara T., Gallart C., Massana P., Nidever D.~L., Cassisi S., et al., 2024, MNRAS, 532, 4272.

\bibitem[\protect\citeauthoryear{Sakowska et al.}{2026}]{Sakowska2026} Sakowska J.~D., Mart{\'\i}nez-Delgado D., Pearson S., Riquel-Castilla F.~J., Starkenburg T.~K., Donatiello G., Deason A., et al., 2026, A\&A, 707, L1.

\bibitem[\protect\citeauthoryear{Sola et al.}{2025}]{Sola2025} Sola E., Duc P.-A., Urbano M., Richards F., Paiement A., B{\'\i}lek M., Y{\i}ld{\i}z M.~K., et al., 2025, MNRAS, 541, 3015.

\bibitem[\protect\citeauthoryear{Starkenburg \& Pearson}{in prep.}]{Starkenburg}
Starkenburg T.~K., Pearson S., in preparation

\bibitem[\protect\citeauthoryear{Torrealba et al.}{2018}]{Torrealba2018} Torrealba G., Belokurov V., Koposov S.~E., Bechtol K., Drlica-Wagner A., Olsen K.~A.~G., Vivas A.~K., et al., 2018, MNRAS, 475, 5085. 

\bibitem[\protect\citeauthoryear{Wheeler et al.}{2015}]{W15} Wheeler C., O{\~n}orbe J., Bullock J.~S., Boylan-Kolchin M., Elbert O.~D., Garrison-Kimmel S., Hopkins P.~F., et al., 2015, MNRAS, 453, 1305. 

\bibitem[\protect\citeauthoryear{White \& Rees}{1978}]{WR78} White S.~D.~M., Rees M.~J., 1978, MNRAS, 183, 341. 


\end{thebibliography}
\end{document}